\documentclass[12pt]{article}
\usepackage{bm, commath}
\usepackage{natbib}
\usepackage{caption}
\usepackage{graphicx}
\usepackage{subcaption}
\usepackage{amsmath, amsfonts, amsthm}
\usepackage{float}
\usepackage{booktabs,siunitx}
\usepackage{url}
\usepackage{multirow}
\usepackage{amssymb}
\usepackage{bm,bbm}
\usepackage{mathrsfs}
\usepackage{multirow}
\usepackage{authblk}
\usepackage{listings}
\allowdisplaybreaks
\usepackage{bbm}
\usepackage{textcomp}
\usepackage{geometry}
\usepackage{authblk}
\usepackage[ruled]{algorithm2e}
\SetKwInput{KwParam}{Parameter}
\SetAlgoCaptionLayout{centerline}

\usepackage{sectsty}
\usepackage[doublespacing]{setspace}
\usepackage{geometry}
\usepackage{hyperref}
\usepackage[utf8]{inputenc}
\usepackage{float}

\newtheorem{theorem}{Theorem}

\newtheorem{remark}{Remark}

\newtheorem{condition}{Condition}

\usepackage{mathtools}

\usepackage{xcolor}
\usepackage{diagbox}

\makeatletter
\renewcommand{\algocf@captiontext}[2]{#1\algocf@typo. \AlCapFnt{}#2} 
\def\@algocf@capt@plain{top}
\renewcommand{\algocf@makecaption}[2]{%
  \addtolength{\hsize}{\algomargin}%
  \sbox\@tempboxa{\algocf@captiontext{#1}{#2}}%
  \ifdim\wd\@tempboxa >\hsize
  \hskip .5\algomargin%
  \parbox[t]{\hsize}{\algocf@captiontext{#1}{#2}}
  \else%
  \global\@minipagefalse%
  \hbox to\hsize{\box\@tempboxa}
  \fi%
  \addtolength{\hsize}{-\algomargin}%
}
\makeatother

\def\E{\mathbbm{E}}
\def\R{\mathbbm{R}}
\def\bz{\bm{z}}

\def\gam{\gamma}
\def\eps{\epsilon}

\def\sig{\sigma}
\def\P{\mathbbm{P}}
\def\V{\mathcal{V}}

\def\S{\mathcal{S}}
\def\C{\mathcal{C}}

\def\B{\mathcal{B}}

\def\bgam{\bm{\gamma}}
\def\bpi{\bm{\pi}}

\def\bgam{\bm{\gamma}}

\def\j{[j]}
\def\k{[k]}

\DeclareMathOperator*{\argmax}{arg\,max}

\newcommand{\keywords}[1]{\textbf{Keywords}:#1}
\sectionfont{\bfseries\large\sffamily}%

\subsectionfont{\bfseries\sffamily\normalsize}%
\usepackage{xr}
\title{Leveraging Local Distributions in Mendelian Randomization: Uncertain Opinions are Invalid}
\author{Ziya Xu and Sai Li\thanks{Corresponding to: saili@ruc.edu.cn}}
\affil{Institute of Statistics and Big Data, Renmin University of China, Beijing, China}
\date{}
\begin{document}

\maketitle

\begin{abstract}
Mendelian randomization (MR) considers using genetic variants as instrumental variables (IVs) to infer causal effects in observational studies. However, the validity of causal inference in MR can be compromised when the IVs are potentially invalid. In this work, we propose a new method, MR-Local, to infer the causal effect in the existence of possibly invalid IVs. By leveraging the distribution of ratio estimates around the true causal effect, MR-Local selects the cluster of ratio estimates with the least uncertainty and performs causal inference within it. We establish the asymptotic normality of our estimator in the two-sample summary-data setting under either the plurality rule or the balanced pleiotropy assumption. Extensive simulations and analyses of real datasets demonstrate the reliability of our approach.
\end{abstract}
\keywords{ Causal inference; Instrumental variable; Mendelian randomization; Pleiotropy. }

\section{Introduction}

The instrumental variable (IV) approach is widely used to infer causal effects in the existence of unmeasured confounders. It relies on the valid IV assumption that instruments only affect the outcome through the exposure of interest. In epidemiology and biological studies, genetic variants are often utilized as IVs to detect causal relationships between phenotypes. Such causal studies are known as Mendelian randomization (MR) and gain popularity in various disciplines \citep{Davey03}. However, the statistical foundations of MR are still evolving due to concerns regarding the potential invalidity of genetic instruments.

The large availability of genome-wide association studies (GWAS) has made genetic variants, particularly single nucleotide polymorphisms (SNPs), a popular choice of IVs. However, the exclusion restriction assumption, a key assumption in conventional IV methods, may not be credible when using genetic instruments. Many genetic variants exhibit pleiotropic effects, meaning that they can affect multiple phenotypes simultaneously \citep{DaveySmith14}. Hence, a variant can affect the outcome through more than one pathway, violating the exclusion restriction assumption. Moreover, the effects and functions of most SNPs are still largely unknown for many biological traits. Researchers are likely to incorporate some invalid IVs in a causal study, which poses a statistical challenge to deal with the possible invalidness given a large number of candidate IVs.

\subsection{Model set-up}
We first specify the potential outcome model with exposure $d_i\in\R$, outcome $y_i\in\R$, and a set of candidate IVs $\bz_i\in\R^p$, $i=1,\dots,n$. We consider the additive linear, constant effects model \citep{holland1988causal,small2007sensitivity}, which is
\begin{align*}
  y_i^{(d,\bz)}-y_i^{(d',\bz')}=(d-d')\beta+(\bz-\bz')^T\bm\eta, ~~\E[ y_i^{(0,\bm{0})}|\bz_i]=\bz_i^T\bm\kappa,
\end{align*}
where  $\beta\in\R$ denotes the causal effect of the exposure on the outcome,  $\bm\eta\in\R^p$ denotes the direct effect of IVs on the outcome, and $\bm\kappa\in\R^p$ denotes the effect of IVs on the outcome through unmeasured confounders. Let $u_i=y_i^{(0,\bm{0})}-\E[ y_i^{(0,\bm{0})}|\bz_i]$.
It gives that
\begin{align*}
    y_i^{(d,\bz)}=d\beta+\bz^T(\bm\kappa+\bm\eta)+u_i,~\E[u_i|\bz_i]=0.
\end{align*}
Let $\bpi=\bm\kappa+\bm\eta\in\R^p$.
The potential outcome model gives the following model for the observed data. For $i=1,\dots,n$,
\begin{align}
\label{eq-y}
    y_i=d_i\beta+\bz_i^T\bpi+u_i,~\E[u_i|\bz_i]=0.
\end{align}
For the exposure $d_i$, we fit a linear working model as
\begin{align}
\label{eq-d}
 d_i=\bz_i^T\bgam_D+v_i,~\text{where}~\bgam_D=\E^{-1}[\bz_i\bz_i^T]\E[\bz_id_i]~\text{and}~\E[v_i\bz_i]=0.
\end{align}
Plugging (\ref{eq-d}) into (\ref{eq-y}), we arrive at the reduced-form representation:
\begin{align}
\label{eq-reduce}
    y_i=\bz_i^T\bgam_Y+\beta v_i+u_i,
\end{align}
where $\E[\bz_i(v_i,u_i)]=\bm{0}$ and
\begin{align}
\label{eq-Gam}
    \gam_{Y,j}=\gam_{D,j}\beta+\pi_{j},~~j=1,\dots,p.
\end{align}
In Equation (\ref{eq-Gam}), $\gam_{Y,j}$ represents the total effect of the $j$-th IV on the outcome, which can be decomposed into two components: the effect through the causal pathway, $\gam_{D,j}\beta$, and the invalid (pleiotropic) effect, $\pi_j$. We define the set of relevant IVs as $\S=\{1\leq j\leq p:\gam_{D,j}\neq 0\}$ and the set of valid IVs as $\V=\{1\leq j\leq p: \gam_{D,j}\neq 0,\pi_j=0\}$. Under conventional IV assumptions, $\V$ is a known (nonempty) set. State-of-the-art methods, such as two-stage least squares \citep{basmann1957TSLS} and inverse-variance weighting \citep{Burgess:2013aa}, can be applied to infer the causal effect. In contrast, we consider the presence of potentially invalid IVs, allowing $\V$ and $\S$ to be unknown a priori.

\subsection{Existing identifiability assumptions}
\label{sec1-2}
Regarding models (\ref{eq-y}) and (\ref{eq-d}), recent research has focused on estimating the causal effect in the presence of invalid IVs. One class of identifiability conditions is based on the prior distribution of $\bpi$. For example, \citet{Bowden15} propose the InSIDE assumption which allows for non-zero invalid effects $\bpi$ provided that the vector $\bpi$ has to be uncorrelated with the instrument strength vector $\bgam_{D}$. On the other hand, \citet{zhao2018statistical} consider the balanced pleiotropy assumption which allows for Gaussian invalid effects with zero mean. An alternative identifiability condition, proposed by \cite{Kole15}, assumes that $\bpi$ and $\bgam_{D}$ are orthogonal, which is equivalent to the InSIDE assumption under the balanced pleiotropy assumption. The above assumptions all relax the standard IV assumptions. However, they are not practically verifiable and may be violated due to the existence of either correlated pleiotropy, where SNPs affect the outcome through confounders \citep{morrison2020ng}, or directional pleiotropy, where $\bpi$ has a non-zero mean.

Another class of identifiability conditions is based on the sparsity  of $\bpi$. Define $\beta^{\j}={\gamma_{Y,j}}/{\gamma_{D,j}}$ for $j\in\S$, also known as the Wald ratio based on the $j$-th IV.  The majority rule \citep{Bowden16,Kang16,Windmeijer2019} assumes that at least half of $\pi_j$'s are zero for $j\in \S$, implying that the median of $\{\beta^{[j]}\}_{j\in\S}$ gives the true causal effect $\beta$. 
A weaker assumption than the majority rule, the plurality rule \citep{TSHT,Wind21} or the ZEMPA assumption \citep{Hartwig:2017aa}, identifies $\beta$ as the mode of $\{\beta^{[j]}\}_{j\in\S}$. Specifically, let $C_j=\{k\in \S:\beta^{\k}=\beta^{\j}\}$ denote the IVs that have same Wald ratio as the $j$-th IV. The plurality rule assumes that $C_{j^*}=\V$ for $j^*=\argmax_{j\in \S}|C_j|$  \citep{TSHT}. In other words, if we cluster the relevant IVs according to their Wald ratios, then the largest cluster must be formed by valid IVs. Hence, the IVs in the largest cluster can be used to identify the true causal effect.  Alternatively, we can define the distribution function of these Wald ratios as $d_{\beta}(t)=\sum_{j\in \S}\mathbbm{1}(\beta^{\j}=t)/|\S|$ for $t\in\R$, so the plurality rule implies that the causal effect can be identified as the mode of $d_{\beta}(t)$. The ZEMPA assumption is equivalent to the plurality rule, and we use the term ``plurality rule'' to refer to this category of assumptions. Although weaker than the conventional IV assumptions, neither the majority rule nor the plurality rule are verifiable based on data.

Many other methods have emerged for estimating causal effects in the presence of invalid IVs. For example, \citet{qi2019mendelian}, \citet{hu2022mendelian}, and \citet{morrison2020ng} propose Gaussian mixture models to account for directional or correlated pleiotropy. However, these methods lack theoretical guarantees, which can make their causal inference results less reliable. Another approach, introduced by \citet{Verbanck:2018aa}, is the MR-PRESSO test, designed to detect horizontal pleiotropy and pleiotropic outlier variants.  Additionally, \citet{sun2022selective} leverage machine learning methods to model the nuisance
parameters including pleiotropic effects. While these new methods relax conventional IV assumptions, they are hard to verify in a data-dependent way. Thus, it is crucial to develop causal inference methods that rely on even weaker assumptions, enhancing the reliability of causal inference in real-world applications.
\subsection{Rationale and our findings}\label{sec-rationale}

In this work, we propose a method to infer the causal effect $\beta$ in more general scenarios than the plurality rule and the balanced pleiotropy assumption. Our rationale can be illustrated as follows. Consider that the invalid effects $\pi_j$ are \textit{i.i.d.} generated from the following model.
\begin{align}
\label{eq1-1}
  \pi_j=0~\text{if}~j\in \V~\text{ and }~\pi_j\sim N(\mu_j,\sig_{\pi}^2) ~\text{if}~j\notin \V.
\end{align}
Model (\ref{eq1-1}) applies to various scenarios with different values of $|\V|$ and $\mu_{j}$. For instance, $|\V|\geq0.5p$ indicates the majority rule holding and $\mu_j=0$ represents the balanced pleiotropy.

\begin{figure}[!h]
	\centering
	\includegraphics[height=4cm,width=7cm]{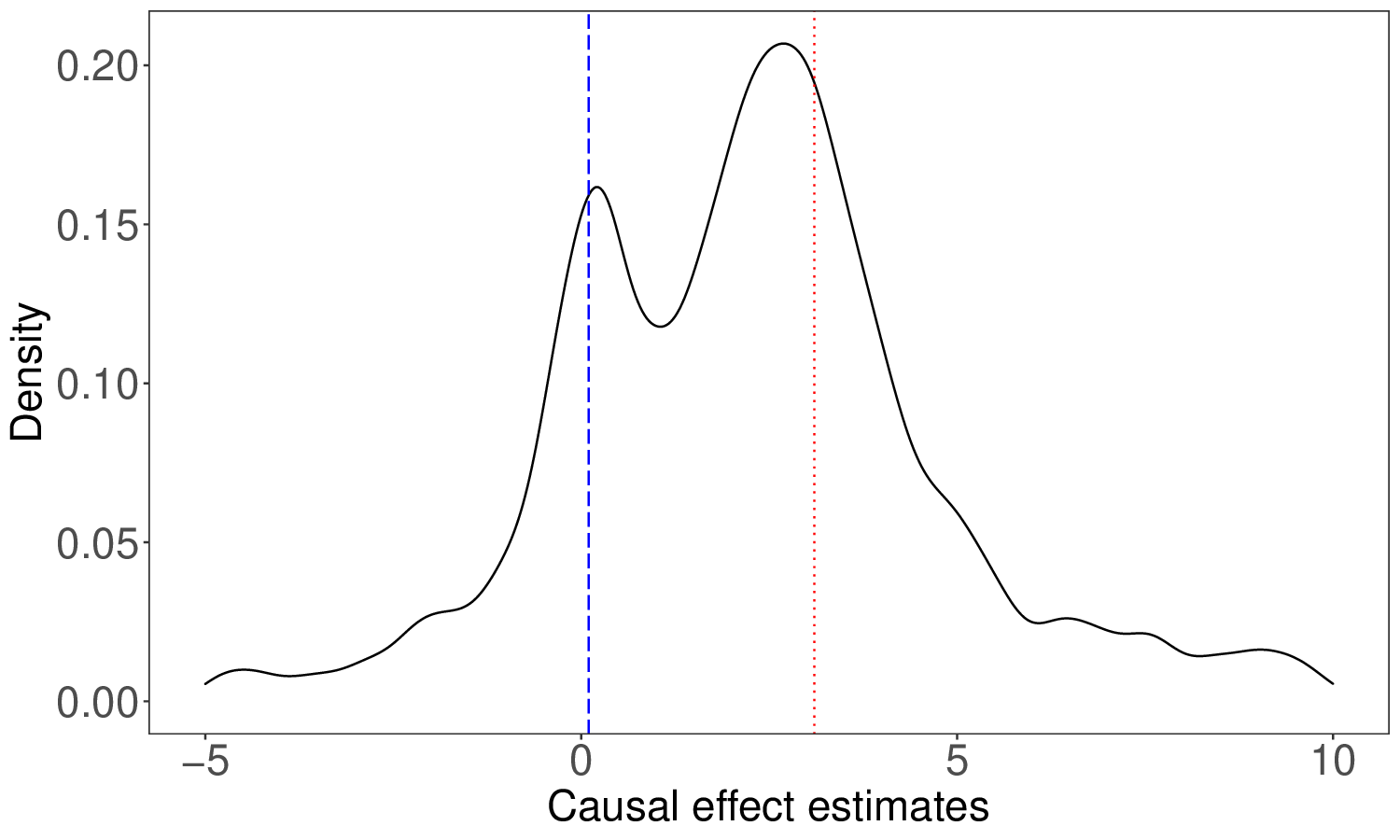}
	\caption{Density plot of the causal effect estimates given by 2000 IVs among which 350 are valid. We set $\beta=0.1$ and $\pi_j=3\gam_{D,j}$ for $j\in\V^c$. The true causal effect is 0.1(blue) and the model of causal effect given by invalid IVs is 3.1(red).}
	\label{fig-intuition1}
\end{figure}

 In practice, we obtain ratio estimates $\hat{\beta}^{[j]}$, the empirical version of ${\beta}^{[j]}$, where $\hat{\beta}^{[j]}={\beta}^{[j]}+\hat{\eps}_j$ contains noise $\hat{\eps}_j$, $j=1,\dots,p$. Assume for now that $\{\hat{\eps}_j\}_{j\leq p}$ are asymptotically normal with variance $\xi^2$ independently. Then, the ratio estimates $\hat\beta^{\j}$, $j\in\S$, satisfy that
\begin{align}
	\label{eq1-2}
\frac{\hat{\beta}^{\j}-\beta}{\xi}\stackrel{D}{\rightarrow}N(0,1)~\text{if}~j\in \V,~~\frac{\hat{\beta}^{\j}-(\beta+{\mu_j}/{\gam_{D,j}})}{\sqrt{\xi^2+{\sig_{\pi}^2}/{\gam_{D,j}^2}}}\stackrel{D}{\rightarrow}N(0,1)~\text{if}~j\notin \V.
\end{align}
If $\mu_j/\gam_{D,j}=\mu_{bias}$ holds for any $j\in\V^c$ and some nonzero constant $\mu_{bias}$, then, as shown in (\ref{eq1-2}), the density of $\{\hat\beta^{\j}\}_{j\leq p}$ has two peaks. One is at $\beta$, the center of valid ratio estimates, while the other is at $\beta + \mu_{bias}$, the center of invalid ratio estimates.
Figure \ref{fig-intuition1} provides an illustrative example where $\beta=0.1$ and $\mu_{bias}=3$.  Methods based on the plurality rule identify $\beta$ as the mode of ratio estimates, which corresponds to the highest peak in the density plot of $\{\hat\beta^{\j}\}_{j\leq p}$. These methods fail to identify $\beta$ in this example since the highest peak is $\beta+\mu_{bias}$.
 Nevertheless, we can identify $\beta$ as the sharpest peak provided that $\sig_{\pi}^2>0$. This idea bridges the plurality rule and the balanced pleiotropy. If the plurality rule holds, the highest peak is also the sharpest. In the case of balanced pleiotropy, the density of ratio estimates only has a single peak at $\beta$ since $\mu_j=0$, which is also the sharpest. Moreover, we can extend this idea to scenarios involving non-Gaussian invalid effects beyond the model (\ref{eq1-1}), provided that the variance of invalid effects is non-zero.

 Based on this rationale, we propose a method to identify the causal effect as the sharpest peak of the distribution of ratio estimates. We first calculates the uncertainty of the estimates in the neighborhood of each potential peak and then performs causal inference within the neighborhood with the least uncertainty, corresponding to the sharpest peak. We also establish the proposed estimator's asymptotic normality in any of the three scenarios outlined in Section \ref{sec3}. In extensive experiments, including GWAS-simulated studies and analyses of two-sample GWAS summary data, our approach consistently exhibits robust performance.

\subsection{Notations and organization}
We introduce some notations. Let $a_n$ and $b_n$ be two sequences of real numbers indexed by $n$. We define $a_n=O(b_n)$ if there exists a constant $c>0$ such that $|a_n|\leq cb_n$ for all $n$, $a_n=o(b_n)$ if $a_n/b_n\rightarrow 0$ as $n\rightarrow \infty$, $a_n\gg b_n$ if $a_n/b_n\rightarrow \infty$ as $n\rightarrow \infty$, and $a_n\asymp b_n$ if there exists a constant $c>0$ such that $c^{-1}b_n\leq |a_n|\leq cb_n$ for all $n$. We write $\xrightarrow{\P}$ to denote convergence in probability, and $\xrightarrow{D}$ to denote convergence in distribution. For a random variable $X$, we denote the expectation of $X$ as $\E(X)$ and the variance of $X$ as $\mathrm{Var}(X)$. If $X$ follows a standard normal distribution, $\Phi(\cdot)$ denotes the cumulative distribution function and $\phi(\cdot)$ denotes the density function. For a set $A$, we denote the complement of set $A$ as $A^c$ and the number of elements in set $A$ as $|A|$.

In the rest of this paper, we introduce the proposed method in Section \ref{sec2}. Theoretical guarantees are provided in Section \ref{sec3}. Numerical results based on simulated GWAS data are conducted in Section \ref{sec-simu}. The proposed method is applied to real studies in Section \ref{sec-data}. Section \ref{sec-diss} concludes the paper with a discussion and some extensions.

\section{Local-distribution based method for MR}
\label{sec2}
In this section, we formalize the idea of using the local distribution to estimate $\beta$. We focus on a two-sample MR setting where we obtain $\hat{\bgam}_D$ as an estimate of $\bgam_D$ from one GWAS and $\hat{\bgam}_Y$ as an estimate of $\bgam_Y$ from another independent GWAS. Their standard errors $\sig_{D,j}$ and $\sig_{Y,j}$, $j=1,\dots,p$ are also available in the corresponding GWAS. While the GWAS statistics are marginal effects, under certain conditions \citep{zhao2018statistical,ye2021debiased}, they are element-wise asymptotically normal such that
\begin{align*}
     \frac{\hat{\gam}_{D,j}-\gam_{D,j}}{\sig_{D,j}}\xrightarrow{D} N(0,1)~~\text{and}~\frac{\hat{\gam}_{Y,j}-\gam_{Y,j}}{\sig_{Y,j}}\xrightarrow{D} N(0,1), ~j=1,\dots,p,
\end{align*}
as $(n_D,n_Y)\rightarrow\infty$. Formal assumptions on the empirical estimates are given in Condition \ref{cond1}. 

We first introduce the key device, the so-called local distribution, for the proposed method. Consider a statistic $\hat{z}_j(b)$ as a function of $b$, which is
\begin{align}\label{def-zj}
    \hat{z}_j(b)= \frac{\hat{\gam}_{Y,j}-b\hat{\gam}_{D,j}}{\sqrt{\sig_{Y,j}^2+b^2\sig_{D,j}^2}},~j=1,\dots,p.
\end{align}
In fact, $\hat z_j(b)$ is standardized $\hat\beta^{\j}-b$ for $\hat{\beta}^{[j]}=\hat\gam_{Y,j}/\hat\gam_{D,j}$. For $\delta_n>0$, we define the local distribution of $\hat z_j(b)$ as
\begin{equation}\label{def-local}
	\mathcal{F}(b,\delta_n)=\P(\hat{z}_j(b)\leq t\mid|\hat\gam_{Y,j}-b\hat\gam_{D,j}|\leq \delta_n).
\end{equation}

We utilize the special property of $\mathcal{F}(\beta,\delta_n)$ to identify $\beta$. We see that $\hat{z}_j(\beta)\xrightarrow{D} N(0,1)$ for $j\in\V$. On the other hand, $\{\hat{\beta}^{[j]}\}_{j\in\V}$ are expected to be closer to $\beta$ than $\{\hat{\beta}^{[j]}\}_{j\in\V^c}$. Thus, for a small enough $\delta_n$, the cluster $\{j:|\hat\gam_{Y,j}-\beta\hat\gam_{D,j}|\leq \delta_n\}$ only contains valid IVs with high probability. In this case,  $\mathcal{F}(\beta,\delta_n)$ is approximately a truncated standard normal distribution:
 \begin{align}
 	\mathcal{F}(\beta,\delta_n)
 	&=\P\left(z\leq t\mid|z|\leq \frac{\delta_n}{\sqrt{\sig_{Y,j}^2+\beta^2\sig_{D,j}^2}}\right)+o(1),\label{local-dist}
 \end{align}
 where $z$ is a standard normal variable. We can identify $\beta$ from a set of candidate values $\{b_j\}_{j\leq p}$ based on (\ref{local-dist}). Various methods can be used to measure the difference between $\mathcal{F}(b_j,\delta_n)$ and the distribution in (\ref{local-dist}) such as the Kullback–Leibler divergence, KS-type statistics, and moment matching.  In the next subsection, we use the second moment of $\mathcal{F}(\beta,\delta_n)$ to estimate $\beta$. We discuss other possible methods in Section \ref{sec-diss}.


\subsection{Proposed algorithm}
\label{sec2-1}

Based on the idea formulated previously, we develop a three-step method to estimate the causal effect.

In the first step, we propose a statistic $\widehat Q(b)$ to test whether the local distribution (\ref{local-dist}) holds around $b$. 
We define a cluster of IVs that have ratio estimates around $b$ as 
\begin{align}
	\label{eq-Cj}
	\widehat{\C}(b)=\left\{1\leq j\leq p: |\hat\gam_{Y,j}-b\hat\gam_{D,j}|\leq \tau_0\sqrt{\sig_{Y,j}^2+b^2\sig_{D,j}^2}\right\}.
\end{align}
The tuning parameter $\tau_0$ in (\ref{eq-Cj}) controls the ``bandwidth''. Its form is studied both theoretically in Section \ref{sec3} and numerically in the Section \ref{sec-simu}. We see that the distribution of $\hat z_j(b)$ for $j\in\widehat{\C}(b)$ is $\mathcal{F}(b,\tau_0\sqrt{\sig_{Y,j}^2+b^2\sig_{D,j}^2})$ as defined in (\ref{def-local}). Consequently, the distribution of $\hat z_j(\beta)$ for $j\in\widehat{\C}(\beta)$ is asymptotically a standard normal distribution truncated at $\pm\tau_0$ as given by (\ref{local-dist}), provided that $\widehat{\C}(\beta)$ only contains valid IVs. Based on this result, we construct the test statistic $\widehat{Q}(b)$ as the standardized second moment of $\{\hat z_j(b)\}_{j\in\widehat{\C}(b)}$. Formally, define
\begin{align}
	&\widehat{Q}(b)=\frac{1}{|\widehat{\C}(b)|}\sum_{j\in \widehat{\C}(b)} \frac{\hat{z}_j^2(b)}{g(\tau_0)}\text{ and }g(\tau_0)=1-\frac{2\tau_0\phi(\tau_0)}{\Phi(\tau_0)-\Phi(-\tau_0)},
	\label{Qhat-j}
\end{align}
where $g(\tau_0)$ is the theoretical variance of a standard normal variable truncated at $\pm \tau_0$. We use the statistic $\widehat{Q}(b)$ to indicate the uncertainty, or equivalently, the sharpness locally around $b$.

In the second step, we perform an uncertainty test based on $\widehat{Q}(b)$. Specifically,  we know that $\E[\widehat{Q}(\beta)|\widehat{\C}(\beta)=\V]=1$. Hence, we search for $b$ whose $\widehat{Q}(b)$ is close to 1. A candidate set for searching can be constructed as follows. Suppose that $\beta$ falls in the interval $[-C_{\beta},C_{\beta}]$ for some positive constant $C_{\beta}$. We define the candidate values $\{b_j\}_{j\leq p}$, ranging from $-C_{\beta}$ to $C_{\beta}$ with a step of $2C_{\beta}/p$. Given the set $\{b_j\}_{j\leq p}$, we define the set of values that pass the uncertainty test as
\begin{equation}
	\label{eq-Jhat}
	\widehat{\B}=\left\{b_j: |\widehat{Q}(b_j)-1|\leq \sig_Q\sqrt{\frac{\log p}{|\widehat{\C}(b_j)|}}+\frac{1}{\log p},|\widehat{\C}(b_j)|\geq \sqrt{p},1\leq j\leq p\right\},
\end{equation}
where $\sig_Q^2={g^{-2}\left(\tau_0\right)}\left({g\left(\tau_0\right)\left(3+\tau_0^2\right)-\tau_0^2}\right)-1.$ In (\ref{eq-Jhat}), the threshold for $\widehat{Q}(b_j)$ is determined by the limiting distribution of $\widehat{Q}(b_j)$ when $b_j=\beta$ and also includes a possible bias term of the GWAS statistics. Furthermore, we exclude the value $b_j$ with a cluster size smaller than $\sqrt{p}$, aligning with our assumption that the number of valid IVs is at least $\sqrt{p}.$ Within the set $\widehat{\B}$, we find $\hat{b}=\argmax_{b\in \widehat{\B}}|\widehat{\C}(b)|$, which corresponds to the highest peak or the mode.

In the last step, we adopt the debiased inverse-variance weighting (dIVW) estimator \citep{ye2021debiased} with the IVs in $\widehat{\C}(\hat b)$ as in (\ref{def-dIVW}) due to its robustness to weak IVs. 
We also construct the confidence interval based on the limiting distribution of the final estimator.

If $\widehat{\B}$ is empty, it indicates that all the clusters have large uncertainty, which can occur in the case of balanced pleiotropy. In this scenario, we employ the dIVW estimator with all candidate IVs to infer $\beta$ and adjust its variance.

We refer to our method as "MR-Local" and present the formal algorithm in Algorithm \ref{alg1}.
\begin{remark}
	The format of (\ref{Qhat-j}) is closely related to Cochran's $Q$-statistic \citep{cochran1954combination}, which can be used to quantify heterogeneity within a vector of statistics. \citet{Bowden:2019aa} estimate $\beta$ by a weighted average of $\hat{\beta}^{\j}$, $j=1,\dots,p$, where the weights are selected to minimize the overall Cochran's $Q$-statistic. Here we use (\ref{Qhat-j}) to detect the set of valid IVs instead of direct estimation. 
\end{remark}

\begin{remark}
	In addition to the balanced pleiotropy, an empty set $\widehat{\B}$ in (\ref{eq-Jhat}) can arise in various scenarios, including the case where all the instruments exhibit directional pleiotropy. To determine the cause,  one can compare  dIVW in (\ref{def-dIVW}) with MR-Raps \citep{zhao2018statistical} when $\widehat{\B}$ is empty. If the estimates are similar, it suggests that the balanced pleiotropy can be the cause. 
\end{remark}

\begin{algorithm}[H]
	\SetKwInput{Input}{Input}
	\SetKwInput{Output}{Output}
	{
		\Input{$\{\hat{\gam}_{Y,j},~\sig_{Y,j}\}_{j=1}^p$, $\{\hat{\gam}_{D,j}, \sig_{D,j}\}_{j=1}^p$, the range of true causal effect $[-C_{\beta},C_{\beta}]$, and a tuning parameter $\tau_0$.}
		
		\underline{Step 1.} Compute the test statistics indicating the uncertainty.
		
		\For{ $j=1,\dots,p$}{
			let $b_j=-C_{\beta}+2C_{\beta}j/p$. Construct clusters $\widehat{\C}(b_j)$ defined in (\ref{eq-Cj}). Based on each $\widehat{\C}(b_j)$, compute the  test statistic $\widehat{Q}(b_j)$ defined in (\ref{Qhat-j}).
		}

		\underline{Step 2.} Conduct the uncertainty test. 
		Calculate the set $\widehat{\B}$ defined in (\ref{eq-Jhat}).  If $\widehat{\B}$ is nonempty, $\hat{b}=\argmax_{b\in \widehat{\B}} |\widehat{\C}(b)|$. If $\widehat{\B}$ is empty, define $\widehat{\C}(\hat b)=\{1,\cdots,p\}$.
		
		\underline{Step 3.} Estimate and infer the causal effect $\beta$. Apply the dIVW estimator in the cluster $\widehat{\C}(\hat b)$ to obtain
		\begin{align}\label{def-dIVW}
			\hat{\beta}_{\mathrm{dIVW}}=\frac{\sum_{j\in\widehat{\C}(\hat b)}\sig_{Y,j}^{-2} \hat\gam_{D,j}\hat\gam_{Y,j}}{\sum_{j\in\widehat{\C}(\hat b)}\sig_{Y,j}^{-2}\left(\hat\gam_{D,j}^2-\sig_{D,j}^2\right)}.
		\end{align}
		
		The $(1-\alpha)\times 100\%$ two-sided confidence interval for $\beta$ is $[ \hat{\beta}_{\mathrm{dIVW}}-z_{\alpha/2}\hat{\sig}_{\beta}, \hat{\beta}_{\mathrm{dIVW}}+z_{\alpha/2}\hat{\sig}_{\beta}]$, where $\hat\sig_{\beta}$ is defined in (\ref{def-sigmahat}) when $ \widehat{\B}$ is nonempty, is defined in (\ref{def-balance sig}) when $ \widehat{\B}$ is empty. 
		
		\caption{MR-Local: Local distribution-based method for MR}
		\label{alg1}}
\end{algorithm}

\begin{remark}
	The majority rule \citep{Bowden16} and plurality rule \citep{TSHT} can be viewed as a vanilla version of Algorithm \ref{alg1} with $\widehat{\B}=\{1,\dots,p\}$. These approaches estimate the causal effect solely based on the largest cluster of IVs. In contrast, MR-Local leverages additional information from the distribution of valid ratio estimates as captured by (\ref{local-dist}).
\end{remark}

\begin{remark}
	Algorithm \ref{alg1} does not require screening for strong IVs compared to \citet{TSHT} and \citet{Windmeijer2019}. This is done by employing two strategies to tackle weak IV bias. First, we use the format $\hat{z}_j(b)$ in (\ref{def-zj}) rather than $\hat\beta^{\j}$, as $\hat\beta^{\j}$ is unstable when $\hat\gam_{D,j}$ approaches zero. Second, we adopt the dIVW estimator. These strategies together ensure that MR-Local is robust to weak IVs.
\end{remark}

\section{Theoretical justifications for Algorithm \ref{alg1}}
\label{sec3} 
In this section, we provide theoretical guarantees for Algorithm \ref{alg1} in each of the following scenarios: (i) the plurality rule, (ii) the balanced pleiotropy, and (iii) a directional pleiotropy case where the plurality rule is violated. We focus on a two-sample MR setting where $\hat{\bgam}_D$ and $\hat{\bgam}_Y$ are generated from two independent GWAS with sample sizes $n_D$ and $n_Y$, respectively. Detailed assumptions are given as follows.


\begin{condition}[Sub-Gaussian noises]
	\label{cond1}
	Assume that
	\[
	{\hat{\gam}_{D,j}-\gam_{D,j}}=\frac{1}{n_D}\sum_{i=1}^{n_D}{\delta^{(D)}_{i,j}}+ {\sig_{D,j}}rem_{D,j}~~\text{and}~~{\hat{\gam}_{Y,j}-\gam_{Y,j}}=\frac{1}{n_Y}\sum_{i=1}^{n_Y}{\delta^{(Y)}_{i,j}}+{\sig_{Y,j}}rem_{Y,j},
	\]
	for $j=1,\cdots,p$,	where $\{\delta^{(D)}_{i,j}\}_{i\leq n_D,j\leq p}$ and $\{\delta^{(Y)}_{i,j}\}_{i\leq n_Y,j\leq p}$ are mutually independent sub-Gaussian variables with zero means and bounded sub-Gaussian norms.
	Suppose that $\sig^2_{D,j}=\sum_{i=1}^{n_D}\mathrm{Var}[\delta^{(D)}_{i,j}]/{n_D^2}$, $\sig^2_{Y,j}=\sum_{i=1}^{n_Y}\mathrm{Var}[\delta^{(Y)}_{i,j}]/{n_Y^2}$, $\max_{j\leq p}|\gamma_{D,j}|=O(p/\sqrt{n_Y\log ^2p})$, and the reminder terms $\{rem_{D,j}\}_{j\leq p}$ and $\{rem_{Y,j}\}_{j\leq p}$ satisfy $\max_{j\leq p}\{|rem_{D,j}|,|rem_{Y,j}|\}=o(1/\sqrt{\log p})$. Moreover, $\min\{n_D, n_Y\}\gg \log^4 p$ and 
	$ {n_Y}/{n_D}=O({p}/{\log p})$.
\end{condition}

Condition \ref{cond1} assumes that the errors in $\hat\gam_{D,j}$ and $\hat\gam_{Y,j}$ are sub-Gaussian variables, allowing for potential bias to account for model misspecification errors and measurement errors. This noise assumption is less restrictive compared to the normality and unbiasedness assumptions made in \citet{zhao2018statistical} and \citet{ye2021debiased}. We assume that the variances $\sig_{D,j}^2$ and $\sig_{Y,j}^2$ are known and decrease with the two GWAS sample sizes $n_D$ and $n_Y$, respectively. Accurate estimates of these variances can be obtained in practice \citep{Bowden15,zhao2018statistical}. The upper bound on the maximum IV strength is typically necessary since $\sum_{ j\leq p}\gam_{D,j}^2=O(1)$ if the exposure's variance exists \citep{zhao2018statistical}.  The assumptions regarding $n_D$, $n_Y$, and $p$ are mild. Specifically, the assumption on $n_Y/n_D$ is weaker than the assumption made in \citet{zhao2018statistical} and \citet{ye2021debiased}, where both papers assume that $n_D\asymp n_Y$.



\subsection{Conclusions under the plurality rule and the balanced pleiotropy assumption}\label{sec-existing}
In this section, we consider two common IV assumptions: the plurality rule and the balanced pleiotropy assumption. We establish the asymptotic normality of our proposed estimator under either assumption.

First, we show the results when the plurality rule holds.
\begin{condition}[The plurality rule holds]\label{cond-PR}
	\begin{itemize}
		\item[(a)] 	Assume that  $|\beta|\leq C_{\beta}$. It holds that $\sup_{b\neq \beta}|\C(b)|\leq c_0|\V|$ for some constant $c_0<1$, where
		\begin{align}\label{def-C(b)}
			\C(b)=\left\{j:|\gam_{Y,j}-b\gam_{D,j}|\leq 2\tau_0\sqrt{\sig_{Y,j}^2+b^2\sig_{D,j}^2}\right\}.
		\end{align}
		
		Moreover,  $\min_{j\in\V^c} |\pi_j|/\sqrt{\sig_{Y,j}^2+C_{\beta}^2\sig_{D,j}^2}> 2\tau_0$ and $|\V|\geq \sqrt{p}$.

		\item[(b)] Denote the average IV strength in a non-empty set $\C$ as $\kappa_{\C}$, where
		\begin{align*}
			\kappa_{\C}=\frac{1}{|\C|} \sum_{j\in \C} \frac{\gamma_{D,j}^2}{\sigma_{D,j}^2}. 
		\end{align*} Suppose that $\max_{j\leq p}\{|rem_{D,j}|,|rem_{Y,j}|\}=o(1/\sqrt{|\V|})$,  the average valid IV strength $\kappa_{\V}\gg n_Dn_Y^{-1}|\V|^{-1}+ |\V|^{-1/2}$, and
		${\max_{ j\in\V}\gam_{D,j}^2}/(\sum_{ j\in\V}\gam_{D,j}^2)=o(1)$.
	\end{itemize}
\end{condition}

Part (a) is a finite sample version of the plurality rule. As the sample sizes grow to infinity,  $\C(\beta^{\j})$ defined in (\ref{def-C(b)}) converges to $C_j$ in \citet{TSHT} and the condition on $|\pi_j|$ converges to $\min_{j\in\V^c}|\pi_j|> 0$. The former two statements combined with $\sup_{b\neq \beta}|\C(b)|\leq c_0|\V|$ is equivalent to the definition of the plurality rule.  The condition that $|\V|\geq \sqrt{p}$ ensures the convergence of the empirical valid IV set, which is usually met when the plurality rule holds. We do not make any assumptions regarding the minimum valid IV strength that allows for the existence of weak IVs, in contrast to \citet{TSHT} and \citet{Wind21}.

In part (b), we assume that the average valid IV strength $\kappa_{\V}$ is not too small to derive the asymptotic normality. To verify Lindeberg's condition, we assume that
the maximum valid IV strength is small compared to the sum of valid IV strength.  If $n_{D}\asymp n_{Y}$ and $|\V|\asymp p$, our condition on $\kappa_{\V}$ simplifies to $\kappa_{\V}\gg p^{-1/2}$, which recovers the assumptions made in \citet{ye2021debiased}.

\begin{theorem}[Asymptotic normality under the plurality rule]
	\label{thm-PR}
	Assume Conditions \ref{cond1} and \ref{cond-PR} (a). Suppose that $\tau_0= c\sqrt{\log p}$, where $c>\sqrt{2}$ and depends on the sub-Gaussian norms in Condition \ref{cond1}. Then, as $(p,n_D,n_Y)\rightarrow\infty$,
	\[
	\P(\widehat{\C}(\hat{b})=\V)\rightarrow 1.
	\]
	
	Further, assume Condition \ref{cond-PR} (b). Then, the dIVW estimator defined in (\ref{def-dIVW}) satisfies that
	\[
	\frac{\hat\beta_{\mathrm{dIVW}}-\beta}{\hat\sig_{\beta}}\stackrel{D}{\rightarrow} N(0,1),
	\]
	where
	\begin{align}\label{def-sigmahat}
		\hat\sig_{\beta}^2=\frac{\sum_{j\in\widehat{\C}(\hat{b})}\sig_{Y,j}^{-4}\left[\hat\gam_{D,j}^2+\hat{\beta}_{ \mathrm{dIVW}}^2 \sig_{D,j}^{2}\left(\hat\gam_{D,j}^2+\sig_{D,j}^2\right)\right]}{\left(\sum_{j\in\widehat{\C}(\hat{b})}\sig_{Y,j}^{-2}\left(\hat\gam_{D,j}^2-\sig_{D,j}^2\right) \right)^2}.
	\end{align}

\end{theorem}
Under conditions of Theorem \ref{thm-PR}, we prove that the selected cluster $\widehat{\C}(\hat b)$ is the set of valid IVs with high probability and the dIVW estimator is asymptotically normal. 
Given the first conclusion of Theorem \ref{thm-PR}, the problem is simplified to infer the causal effect in a valid IV scenario.  The standard IVW estimator assumes the average IV strength to be larger than $p$, which is unrealistic in practice. In contrast, we employ the dIVW estimator proposed by \citet{ye2021debiased}, which ensures asymptotic normality when the average valid IV strength $\kappa_{\V}\gg n_Dn_Y^{-1}|\V|^{-1}+ |\V|^{-1/2}$, given our sub-Gaussian noises assumption.

Next, we show the results under the balanced pleiotropy assumption.
\begin{condition}[The balanced pleiotropy assumption holds]\label{cond-BP}
	Assume that  $|\beta|\leq C_{\beta}$. The invalid effects satisfy that
	\begin{align*}
		{\pi_{j}}\sim_{i.i.d.} N(0,\sig_{\pi}^2)~\text{for}~j\leq p,
	\end{align*}
	where $ \sig_{\pi}\geq c_1 \max_{ j\leq p}\sqrt{\sig_{Y,j}^2+C_{\beta}^2\sig_{D,j}^2}$ for some constant $c_1>0$.
	
	Furthermore, suppose that $\max_{j\leq p}\{|rem_{D,j}|,|rem_{Y,j}|\}=o(1/\sqrt{p})$, the average IV strength  $\kappa=\kappa_{\{1,\cdots,p\}}\gg n_Dn_Y^{-1}p^{-1}+p^{-1/2}(1+\max_{j\leq p}\sig_{\pi}\sig_{D,j}^{-1})$, and ${\max_{ j\leq p}\gam_{D,j}^2}/(\sum_{ j\leq p}\gam_{D,j}^2)=o(1)$.
\end{condition}
Condition \ref{cond-BP} assumes the balanced pleiotropy assumption studied in  \citet{zhao2018statistical} and \citet{ye2021debiased}. 
The condition on the average IV strength $\kappa$ is needed to derive asymptotic normality. If $n_{D}\asymp n_{Y}$ and $\max_{j\leq p}\sig_{\pi}\sig_{D,j}^{-1}=O(1)$ as implied by \citet{ye2021debiased}, our condition on $\kappa$ simplifies to $\kappa\gg p^{-1/2}$, which recovers the assumption made in \citet{ye2021debiased}.  
The condition on $\sig_{\pi}$ ensures that the invalid IV estimates exhibit large uncertainty, preventing them from passing the uncertainty test as in (\ref{eq-Jhat}). 
\begin{theorem}[Asymptotic normality under the balanced pleiotropy]
	\label{thm-BP}
	Assume Conditions \ref{cond1} and \ref{cond-BP}.
	Suppose that $\tau_0= c\sqrt{\log p}$, where $c>\sqrt{2}$ and depends on the sub-Gaussian norms in Condition \ref{cond1}. Then, as $(p,n_D,n_Y)\rightarrow\infty$,
	\[
	\frac{\hat\beta_{\mathrm{dIVW}}-\beta}{\hat\sig_{\beta}}\stackrel{D}{\rightarrow} N(0,1),
	\]
	where
	\begin{align}\label{def-balance sig}
		\hat\sig_{\beta}^2=\frac{\sum_{j=1}^p\sig_{Y,j}^{-4}\left[(\sig_{Y,j}^{2}+\hat\sig_{\pi}^2)\hat\gam_{D,j}^2+\hat{\beta}_{ \mathrm{dIVW}}^2 \sig_{D,j}^{2}\left(\hat\gam_{D,j}^2+\sig_{D,j}^2\right)\right]}{\left(\sum_{j=1}^p\sig_{Y,j}^{-2}\left(\hat\gam_{D,j}^2-\sig_{D,j}^2\right) \right)^2} 
	\end{align}
	and
	\begin{align}\label{def-balance sigpi}
		\hat\sig_{\pi}^2=\frac{\sum_{j=1}^p\left[(\hat\gam_{Y,j}-\hat\beta_{\mathrm{dIVW}}\hat\gam_{D,j})^2-\sig_{Y,j}^2-\hat\beta_{\mathrm{dIVW}}^2\sig_{D,j}^2\right]\sig_{Y,j}^{-2}}{\sum_{j=1}^p\sig_{Y,j}^{-2}}.
	\end{align}

\end{theorem}

Theorem \ref{thm-BP} justifies the asymptotic normality of our proposed estimator in the balanced pleiotropy case. The variance estimator in (\ref{def-balance sig}) involves $\sig_{\pi}^2$ due to the presence of invalid effects, which is no smaller than the variance estimator in (\ref{def-sigmahat}). We estimate $\sig_{\pi}^2$ by (\ref{def-balance sigpi}) following the approach in \citet{ye2021debiased}.

\begin{remark}\label{re-bp}
	Theorem \ref{thm-BP} requires that $\sig_{\pi}$ is no smaller than the order of $\min\{n_D,n_Y\}^{-1/2}$, while \citet{ye2021debiased} and \citet{zhao2018statistical} prefer a smaller $\sig_{\pi}$ that is not greater than the order of $\min\{n_D,n_Y\}^{-1/2}$.  Notably, our proposed estimator still enjoys asymptotic normality  when $\sig_{\pi}=o(1/\sqrt{n_Y\log p})$ since $\widehat{\C}(\beta)$ could pass the uncertainty test as in (\ref{eq-Jhat}). We provide the formal proof in the supplements. 
\end{remark}

According to Theorems \ref{thm-PR} and \ref{thm-BP}, our proposed causal estimate is consistent and the confidence interval constructed in Algorithm \ref{alg1} achieves the theoretical coverage probability. Our method can handle two scenarios: when the plurality rule holds and when there exists balanced pleiotropy. To the best of our knowledge, no existing approach in the literature can handle both scenarios. Therefore, MR-local provides a robust estimation of the causal effect as it is suitable for more general cases.

\subsection{Further results under another directional pleiotropy assumption}\label{sec-dir}
Besides the two scenarios considered in Section \ref{sec-existing}, we further explore the situation that only a small proportion of valid IVs exists and the pleiotropic effects are not balanced. In this case, methods based on the plurality rule or the balanced pleiotropy assumption can fail. Next, we show that our method still works under mild conditions.
\begin{condition}[A directional pleiotropy assumption]\label{cond-dir}
	Assume that  $|\beta|\leq C_{\beta}$ and $|\V|\geq \sqrt{p}$. The invalid effects satisfy that
	\begin{align*}
		{\pi_{j}}\sim_{i.i.d.} N(\mu_j,\sig_{\pi}^2)~\text{for}~j\in\V^c,
	\end{align*}
	where $\min_{j\in\V^c}|\mu_{j}|/\sqrt{\sig^2_{\pi}+\sig_{Y,j}^2+C_{\beta}^2\sig_{D,j}^2}\geq 2\tau_0$ and $\sig_{\pi}\geq c_1\max_{ j\leq p}\sqrt{\sig_{Y,j}^2+C_{\beta}^2\sig_{D,j}^2}$ for some constant $c_1>1$.	
\end{condition}
In Condition \ref{cond-dir}, we assume that the invalid effect $\pi_j$ follows a Gaussian distribution with mean $\mu_{j}$ and variance $\sig_{\pi}^2$ for $j\in\V^c$. The lower bound on $|\mu_j|$ serves as a separation condition to ensure that Wald ratios based on invalid IVs and those based on valid IVs are separated. Thus, the ratio estimates around the true causal effect are unlikely to be invalid. The lower bound on $\sig_{\pi}$ is similar to the lower bound in Condition \ref{cond-BP}. Condition \ref{cond-dir} covers the scenario where both the balanced pleiotropy assumption and the plurality rule fail. We provide examples and further explanations in the supplements.

\begin{theorem}[Asymptotic normality under directional pleiotropy]\label{thm-dir}
	Assume Conditions \ref{cond1} and \ref{cond-dir}.
	Suppose that $\tau_0= c\sqrt{\log p}$, where $c>\sqrt{2}$ and depends on the sub-Gaussian norms in Condition \ref{cond1}. Then, as $(p,n_D,n_Y)\rightarrow\infty$,
	\[
	\frac{|\widehat{\C}(\hat b)\cap \V^c|}{|\widehat{\C}(\hat b)|}\stackrel{\P}{\rightarrow} 0.
	\]

	Further, define $
	\mathcal{M}=\{j\in\V^c:|\beta^{\j}-\beta|\leq 2\tau_0\rho_{j}(C_{\beta})+2\tau_0\min_{j \in \V}\rho_{j}(C_{\beta})\}
	$, where $\rho_{j}(C_{\beta})=\sqrt{\sig_{Y,j}^2+C_{\beta}^2\sig_{D,j}^2}/|\gam_{D,j}|$ for $\gam_{D,j}\neq 0$. We assume Condition \ref{cond-PR} (b), $n_D\asymp n_Y$, and  $|\mathcal{M}|=o\left(\sqrt{\kappa_{\V}|\V|}/(\kappa_{\mathcal{M}}+\log p)\right)$.
	Then, the dIVW estimator defined in (\ref{def-dIVW}) satisfies that
	\[	\frac{\hat\beta_{\mathrm{dIVW}}-\beta}{\hat\sig_{\beta}}\stackrel{D}{\rightarrow} N(0,1),
	\]
	where $\hat\sig_{\beta}$ is defined in (\ref{def-sigmahat}).
\end{theorem}

Under Conditions \ref{cond1} and \ref{cond-dir}, we demonstrate that the proportion of invalid IVs in the selected cluster $\widehat{\C}(\hat{b})$ converges to zero. Due to the potential large biases introduced by invalid estimates in $\widehat{\C}(\hat{b})$, we need further conditions to establish the asymptotic normality.   We define the misidentified set $\mathcal{M}$ to include invalid IVs with Wald ratios close to $\beta$. When the size of $\mathcal{M}$ is small, our proposed estimator is asymptotic normal under the mentioned conditions. We finally provide other consistent estimators, the primary estimate $\hat{b}$ and the median estimator of $\{\hat\beta^{\j}\}_{j\in\widehat{\C}(\hat{b})}$,  under relaxed IV strength conditions in the supplements.

MR-Egger \citep{Bowden15} is supposed to address this directional pleiotropy scenario by assuming $\mu_j=\mu$ for $j=1,\cdots,p$ under the InSIDE assumption. In contrast, our method does not rely on the InSIDE assumption and provides a viable alternative for handling directional pleiotropy.

\section{Simulated GWAS experiments}
\label{sec-simu}
In this section, we conduct numerical experiments based on simulated GWAS summary statistics. The code for all the methods in comparison is available at \url{https://github.com/saili0103/MR-Local}.

\subsection{Randomly generated effect sizes}
\label{sec-simu-1}
In the first experiment, we simulate the summary statistics with $n_D=n_Y=10^5$ and $p=2000$. Let $h_D=0.1$ represents the heritability of trait $D$. We generate $\gam_{D,j}\sim N(0,h_D/p)$ independently for $j=1,\dots,p$. Therefore, $\sum_{j=1}^p\gam_{D,j}^2\approx h_D$. By the reduced form, we set $\gam_{Y,j}=\gam_{D,j}\beta+\pi_{j}$, where $\beta$ and $\pi_{j}$ vary in different settings.
We generate the standard error of $\gam_{D,j}$, $\sig_{D,j}\sim U[0.8,1]/\sqrt{n_D}$ and the standard error of $\gam_{Y,j}$, $\sig_{Y,j}\sim U[0.8,1]/\sqrt{n_Y}$ independently. The corresponding IV strength $\kappa\approx 6.25$ in this set-up.

\begin{itemize}
	\item[(a)]  $\V$ is a random subset of $\{1,\cdots,p\}$ with $|\V|=0.5p$. Set $\pi_{j}=0$ for $j\in\V$ and $\pi_{j}\sim N(0,0.05/p)+2.5\gam_{D,j}$ for each $j\notin\V$.
	\item[(b)]  $\V$ is a random subset of $\{1,\cdots,p\}$ with $|\V|=0.5p$. Set $\pi_{j}=0$ for $j\in\V$ and $\pi_{j}\sim N(0,0.5/p)+2.5\gam_{D,j}$ for each $j\notin\V$.
	\item[(c)] $\pi_j\sim_{i.i.d.} N(0, 0.1/p)$ for $j=1,\dots,p$.
	\item[(d)]  $\pi_j\sim_{i.i.d.} N(0, 0.05/p)$ for $j=1,\dots,p$.
	\item[(e)]$\V$ is a random subset of $\{1,\cdots,p\}$ with $|\V|= 0.28 p$. Set $\pi_j=0$ for $j\in\V$ and $\pi_j\sim_{i.i.d.} N(0,0.05/p)+2.5\gam_{D,j}$ for each $j\notin\V$.
\end{itemize}
In (a) and (b), the plurality rule holds. The valid ratio estimates form the highest peak in both settings while the invalid ratio estimates are more spread out in (b).
In (c) and (d), the pleiotropic effects are balanced with different $\sig^2_{\pi}$. In (e), valid IVs are relatively few and the invalid ratio estimates form the highest peak. In (a), (b), and (e), a fixed effect $2.5\gam_{D,j}$ is added to each $\pi_{j}$, corresponding to the separation condition in Condition \ref{cond-dir}.

In each experiment, we generate $\hat{\gam}_{D,j}\sim N(\gam_{D,j},\sig_{D,j}^2)$ and $\hat{\gam}_{Y,j}\sim N(\gam_{Y,j},\sig_{Y,j}^2)$ independently.
Our proposal is compared with three other MR methods which are robust to pleiotropy under certain assumptions: MR-MBE \citep{Hartwig:2017aa}, MR-Raps \citep{zhao2018statistical}, and two-stage hard thresholding (TSHT, \citet{TSHT}).
For the robust performance of MR-MBE and TSHT, we first screen out weak IVs and only use the IVs such that $|\hat{\gam}_{D,j}|/\sig_{D,j}\geq \sqrt{2\log p}$. Each setting is replicated based on 500 Monte Carlo experiments.

The proposed MR-Local method incorporates two empirical adjustments for robust inference. First, we apply mild IV strength screening using Algorithm \ref{alg1} with the criterion $|\hat{\gam}_{D,j}|/\sig_{D,j}\geq \tau_0$, where $\tau_0$ is the tuning parameter in (\ref{eq-Cj}). We report results for different choices of $\tau_0$ in the supplements. Second, if $\widehat{B}\neq \emptyset$, the dIVW estimator $\hat{\beta}_{\textup{dIVW}}$ is subject to post-selection effects. Hence, we estimate its standard deviation, $\hat{\sig}_{\beta}$, via bootstrap. Additionally, we introduce another variant of MR-Local called MR-Local+, which incorporates the uncertainty test statistic $\widehat{Q}(b)$ along with a skewness test statistic $\widehat{K}(b)$. The motivation is to refine the selected $\widehat{\B}$ by leveraging the fact that when $b=\beta$, the skewness of $\hat{z}_j(b)$ for $j\in\widehat{C}(b)$ should be close to zero. Further details of MR-Local+ can be found in the supplements.

\begin{figure}[H]
	\centering
	\includegraphics[height=5.1cm,width=0.52\textwidth]{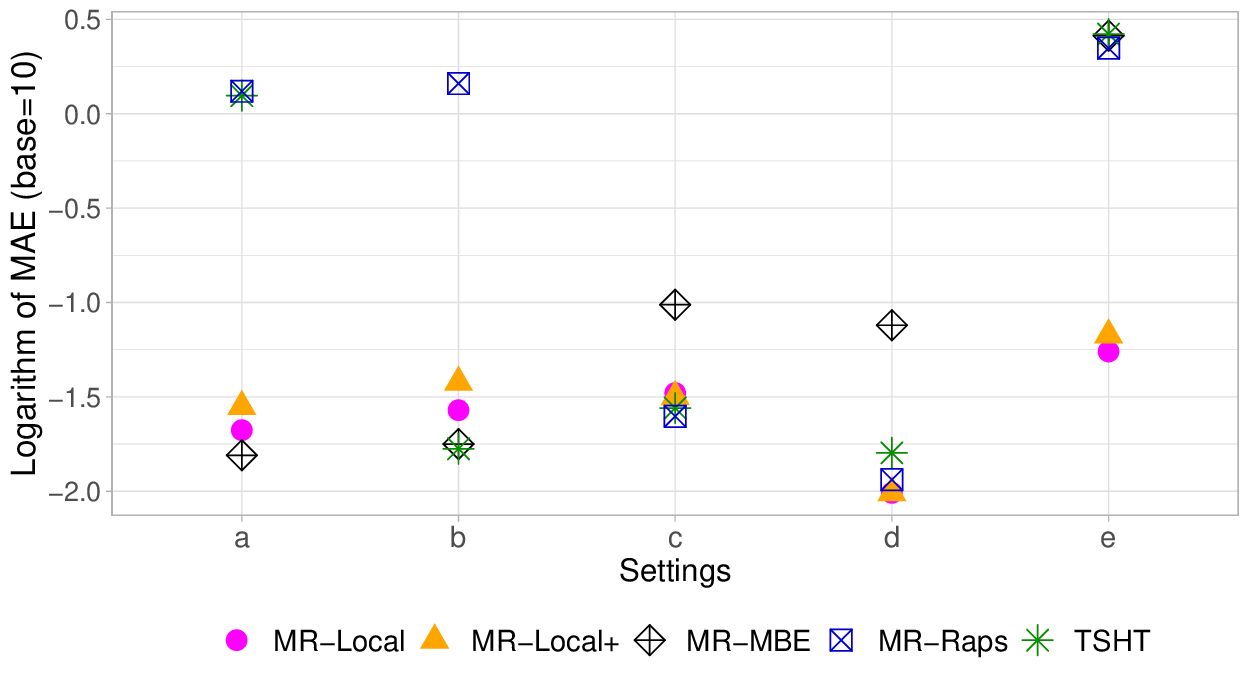}
	\includegraphics[height=5.1cm,width=0.47\textwidth]{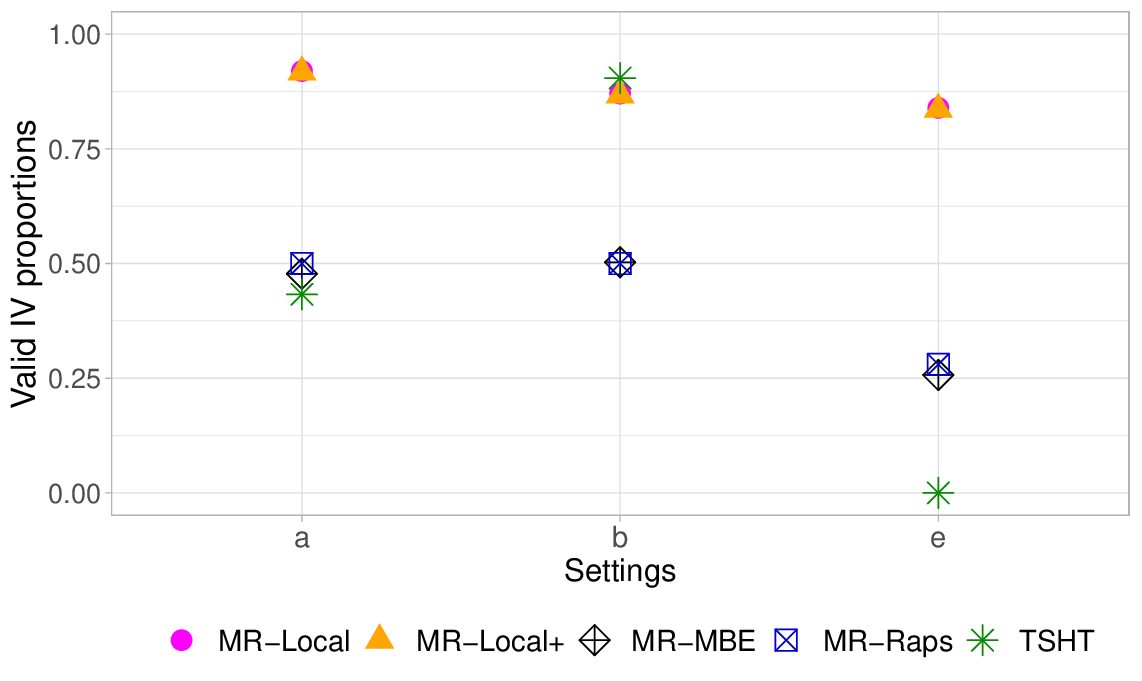}
	\caption{The logarithm of mean absolute errors (MAE, left panel) and the proportion of valid IVs used to estimate the causal effect (right panel) based on MR-Local ($\tau_0=1.6$), MR-Local+($\tau_0=1.6$), MR-Median, and MR-Raps in five settings.}
	\label{fig-mae1}
\end{figure}

In Figure \ref{fig-mae1}, we present the estimation errors of the five methods in settings (a)-(e). Our proposed methods, MR-Local and MR-Local+, have reliable performances across all settings. In contrast, the other methods exhibit significant estimation errors in at least one setting. Specifically, MR-Raps performs poorly when the Gaussian assumption of pleiotropic effects is violated. MR-MBE becomes less accurate in settings (c)-(e) when the proportion of invalid IVs is small. TSHT, designed for the plurality rule, exhibits large errors in settings (a) and (e). In the right panel of the figure, we show the proportion of valid IVs in $\widehat{\mathcal{C}}(\hat{b})$ for our methods. For the other methods, we report the proportion of valid IVs based on their screening steps. Since MR-Raps lacks a screening step and MR-MBE only filters out weak IVs, their proportions of valid IVs are relatively low. TSHT employs a voting step to select valid IVs, which is successful in setting (b) but not in settings (a) and (e), explaining its performance in the estimation errors shown in the left panel. In contrast, our proposal demonstrates a relatively high proportion of valid IVs in $\widehat{\mathcal{C}}(\hat{b})$, highlighting the effectiveness of our screening approach based on the uncertainty test. This, in turn, explains its reliable performance in estimation.

\begin{table}[!htbp]
	\setlength{\tabcolsep}{4.2pt}
	{\renewcommand{\arraystretch}{1.2}%
		\small
		\begin{tabular}{|cc|ccc|ccc|cc|cc|cc|}
			\hline
			Setup&$\beta$&\multicolumn{3}{c|}{MR-Local}&\multicolumn{3}{c|}{MR-Local+} &\multicolumn{2}{c|}{MR-MBE} &\multicolumn{2}{c|}{TSHT} & \multicolumn{2}{c|}{MR-Raps}\\
			\hline
			& &cov. & s.d. &$\emptyset$&cov. & s.d. &$\emptyset$&cov. & s.d.&cov. & s.d.&cov. & s.d.\\
			\hline
			\multirow{2}{*}{a} 
			& 0 & 1.000 & 0.06 & 0.00 & 0.974 & 0.04 & 0.00 & 0.956 & 0.02 & 0.072 & 0.04 & 0.000 & 0.04 \\
			& 0.1  & 0.998 & 0.07 & 0.00 & 0.982 & 0.04 & 0.01 & 0.958 & 0.02 & 0.088 & 0.04 & 0.000 & 0.04 \\
			\hline
			\multirow{2}{*}{b} 
			&0 & 0.992 & 0.05 & 0.00 & 0.962 & 0.03 & 0.02 & 0.948 & 0.03 & 0.656 & 0.02 & 0.000 & 0.04\\
			&0.1& 0.990 & 0.06 & 0.00 & 0.962 & 0.03 & 0.01 & 0.974 & 0.03 & 0.842 & 0.02 & 0.000 & 0.04\\
			\hline
			\multirow{2}{*}{c} &0 & 0.940 & 0.03 & 1.00 & 0.940 & 0.03 & 1.00 & 0.874 & 0.09 & 0.414 & 0.01 & 0.972 & 0.03 \\
			
			&0.1 & 0.914 & 0.03 & 1.00 & 0.916 & 0.03 & 1.00 & 0.900 & 0.09 & 0.504 & 0.01 & 0.976 & 0.03\\
			\hline
			\multirow{2}{*}{d} 
			&0 & 0.996 & 0.02 & 1.00 & 0.998 & 0.02 & 1.00 & 0.948 & 0.08 & 0.682 & 0.01 & 0.996 & 0.02 \\
			&0.1 & 0.998 & 0.02 & 1.00 & 0.998 & 0.02 & 1.00 & 0.956 & 0.08 & 0.722 & 0.01 & 0.992 & 0.02 \\
			\hline
			\multirow{2}{*}{e} &0 & 0.994 & 0.10 & 0.00 & 0.966 & 0.07 & 0.01 & 0.000 & 0.12 & 0.000 & 0.01 & 0.000 & 0.04 \\
			&0.1& 0.986 & 0.12 & 0.01 & 0.970 & 0.07 & 0.01 & 0.000 & 0.12 & 0.000 & 0.01 & 0.000 & 0.04\\
			\hline
		\end{tabular}
	}
	\centering
	\caption{Average coverage probabilities (cov.) and average standard deviation (s.d.) for the 95\% confidence intervals computed via proposed MR-Local($\tau_0=1.6$), MR-Local+($\tau_0=1.6$), MR-MBE, TSHT, and MR-Raps in settings (a) to (e). The column ``$\emptyset$'' indicates the chance of occuring $\{\widehat{\mathcal{B}}=\emptyset\}$.}
	\label{tab-inf1}
\end{table}

In Table \ref{tab-inf1}, we present the inference results for each method. We observe that both MR-Local and MR-Local+ demonstrate coverage probabilities close to the nominal level. Notably, MR-Local+ exhibits higher efficiency with smaller estimated standard errors. As expected, these two methods estimate $\widehat{\mathcal{B}}=\emptyset$ in settings (c) and (d), successfully detecting the presence of balanced pleiotropy. The inference results of MR-Raps are sensitive to the assumption of Gaussian pleiotropy. MR-MBE shows reliable coverage probabilities in settings (a), (b), and (d), but its confidence intervals tend to be longer compared to other methods, resulting in lower efficiency. TSHT exhibits relatively low coverage probabilities, and its intervals are narrow. This is due to the absence of adjustment for post-selection effects in its standard error estimates, whereas the bootstrap method employed in MR-Local, MR-Local+, and MR-MBE partially accounts for selection uncertainty.

\subsection{Simulations based on BMI and SBP GWAS}
\label{sec-simu-2}
In this subsection, we conduct simulations using data from a GWAS on body mass index (BMI) and another GWAS on systolic blood pressure (SBP). The data is available in the R package ``mr.raps'' \citep{mrraps}. Since the original dataset only contains 160 SNPs, we replicate the observed effect vectors and standard error vectors 10 times each to create a new dataset with 1600 SNPs. Specifically, we set $\bgam_{D}$ to be the observed effects from the BMI GWAS after replication, and the standard errors $\{\sig_{D,j}\}_{j\leq  p}$ and $\{\sig_{Y,j}\}_{j\leq p}$, are set to be observed standard errors in the BMI dataset. The pleiotropic effects $\pi_j$ are generated according to each of the following settings. Let $h_D^*=\sum_{j=1}^p\gam_{D,j}^2$.

\begin{itemize}
	\item[(a)]  $\V$ is a random subset of $\{1,\cdots,p\}$ with $|\V|=0.6p$. Set $\pi_{j}=0$ for $j\in\V$ and $\pi_{j}\sim N(0,0.5h_D^*/p)+2.5\gam_{D,j}$ for each $j\notin\V$.
	\item[(b)]  $\V$ is a random subset of $\{1,\cdots,p\}$ with $|\V|=0.4p$. Set $\pi_{j}=0$ for $j\in\V$ and $\pi_{j}$ to be the $j$-th effect on SBP obtained from the data for each $j\notin\V$.
	\item[(c)] $\pi_j\sim_{i.i.d.} N(0, h_D^*/p)$ for $j=1,\dots,p$.
	\item[(d)] $\C$ is a random subset of $\{1,\cdots,p\}$ with $|\C|=0.4p$. Set $\pi_j\sim_{i.i.d.} N(0, h_D^*/p)$ for $j\in \C$ and $\pi_j\sim_{i.i.d.} N(0, 4h_D^*/p)$ for $j\notin \C$.
	\item[(e)]$\V$ is a random subset of $\{1,\cdots,p\}$ with $|\V|= 0.25 p$. Set $\pi_j=0$ for $j\in\V$ and $\pi_j\sim_{i.i.d.} N(0,1)|\gam_{D,j}|+3\gam_{D,j}$ for each $j\notin\V$.
\end{itemize}
In (a) and (b), the plurality rule holds. The pleiotropic effects in (b) are set to be the observed effects, which are different from random Gaussian effects.
In (c) and (d), the pleiotropic effects are balanced. In (d), $\pi_j$ are generated from a Gaussian mixture model centered at zero. Setting (e) corresponds to the situation considered in Section \ref{sec-dir}.
\begin{figure}[H]
	\centering
	\includegraphics[height=5.1cm,width=0.52\textwidth]{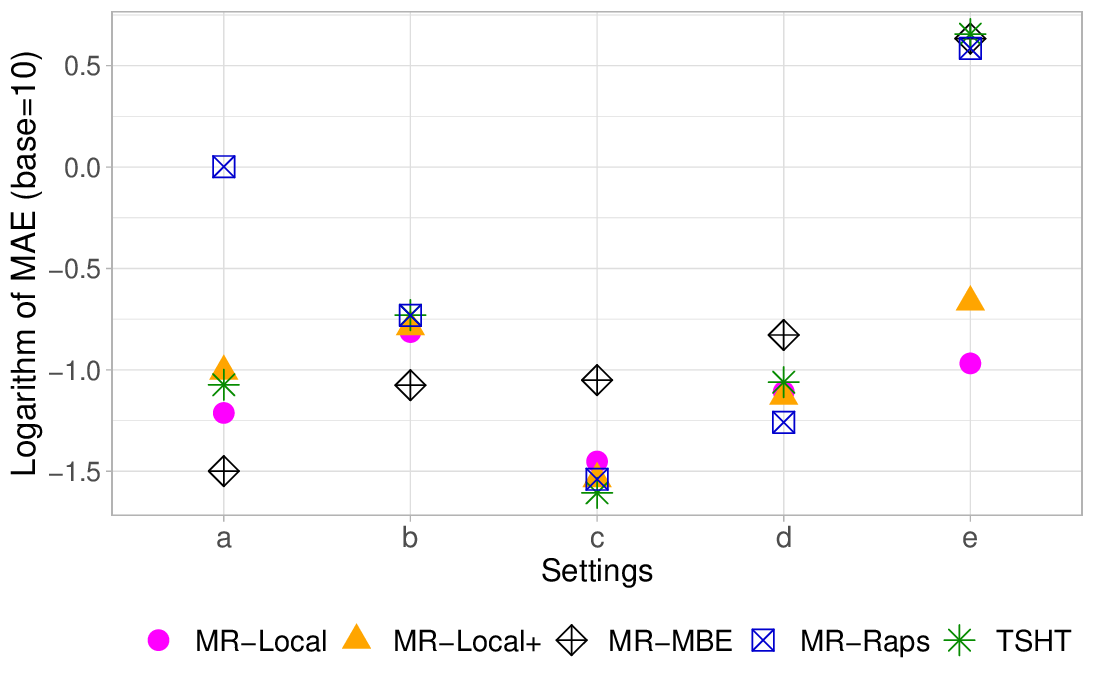}
	\includegraphics[height=5.1cm,width=0.47\textwidth]{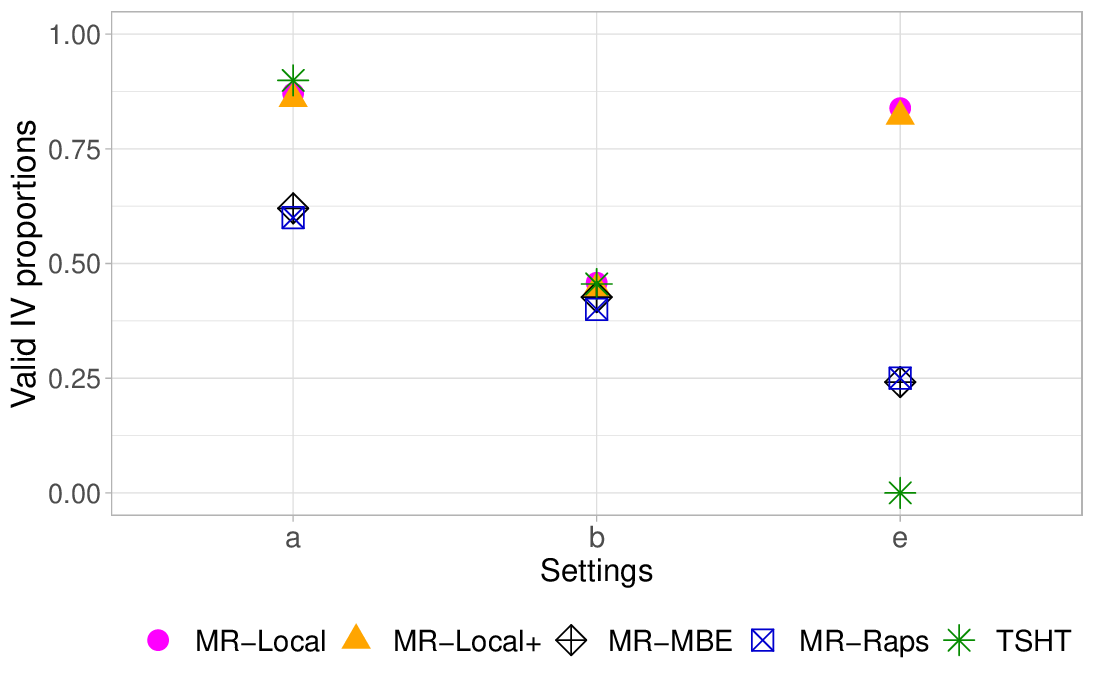}
	\caption{The logarithm of mean absolute errors (MAE, left panel) and the proportion of valid IVs used to estimate the causal effect (right panel) based on MR-Local ($\tau_0=1.6$), MR-Local+($\tau_0=1.6$), MR-Median, and MR-Raps in five settings.}
	\label{fig-mae2}
\end{figure}
From Figure \ref{fig-mae2}, we observe that MR-Local and MR-Local+ perform robustly across various settings. MR-MBE has the smallest estimation errors in settings (a) and (b) but shows relatively large errors in other settings. MR-Raps has the smallest errors in settings (c) and (d) but becomes less accurate when the pleiotropic effects are not Gaussian. TSHT shows significant errors in setting (e), where the plurality rule fails. In Table \ref{tab-inf2}, MR-Local achieves coverage probabilities close to the nominal levels, except for setting (b) where some invalid effects are close to zero, posing a challenge in distinguishing valid IVs from invalid ones. Specifically, MR-Local estimates $\widehat{\mathcal{B}}=\emptyset$ approximately 50\% of the time in setting (b), indicating that no valid peak is identified in half of the cases. In setting (d), where the pleiotropic effects follow a Gaussian mixture distribution, MR-Raps exhibits greater robustness than our proposal.

\begin{table}[!htbp]
	\setlength{\tabcolsep}{4.2pt}
	{\renewcommand{\arraystretch}{1.2}%
		\small
		\begin{tabular}{|cc|ccc|ccc|cc|cc|cc|}
			\hline
			Setup&$\beta$&\multicolumn{3}{c|}{MR-Local}&\multicolumn{3}{c|}{MR-Local+} &\multicolumn{2}{c|}{MR-MBE} &\multicolumn{2}{c|}{TSHT} & \multicolumn{2}{c|}{MR-Raps}\\
			\hline
			& &cov. & s.d. &$\emptyset$&cov. & s.d. &$\emptyset$&cov. & s.d.&cov. & s.d.&cov. & s.d.\\
			\hline
			\multirow{2}{*}{a} 
			& -0.2 & 0.960 & 0.09 & 0.02 & 0.830 & 0.07 & 0.06 & 0.974 & 0.06 & 0.104 & 0.02 & 0.000 & 0.04 \\ 
			&0 & 0.968 & 0.10 & 0.00 & 0.900 & 0.08 & 0.04 & 0.988 & 0.06 & 0.180 & 0.02 & 0.000 & 0.04 \\ 
			\hline
			\multirow{2}{*}{b} 
			&-0.2 & 0.338 & 0.05 & 0.52 & 0.182 & 0.04 & 0.65 & 0.948 & 0.12 & 0.000 & 0.02 & 0.000 & 0.04 \\
			&0& 0.330 & 0.05 & 0.52 & 0.220 & 0.04 & 0.62 & 0.950 & 0.11 & 0.000 & 0.02 & 0.000 & 0.04 \\ 
			\hline
			\multirow{2}{*}{c} &-0.2 & 0.954 & 0.04 & 0.97 & 0.970 & 0.03 & 0.99 & 0.960 & 0.12 & 0.622 & 0.02 & 0.980 & 0.03\\
			
			&0 & 0.970 & 0.04 & 0.97 & 0.982 & 0.04 & 0.98 & 0.964 & 0.12 & 0.796 & 0.02 & 0.984 & 0.03 \\
			\hline
			\multirow{2}{*}{d} 
			&-0.2& 0.920 & 0.05 & 0.96 & 0.938 & 0.05 & 0.98 & 0.810 & 0.14 & 0.088 & 0.02 & 0.966 & 0.05\\
			&0 & 0.898 & 0.05 & 0.98 & 0.904 & 0.05 & 0.99 & 0.856 & 0.14 & 0.174 & 0.02 & 0.946 & 0.05 \\
			\hline
			\multirow{2}{*}{e} &-0.2 & 0.980 & 0.21 & 0.00 & 0.938 & 0.18 & 0.03 & 0.000 & 0.23 & 0.000 & 0.03 & 0.000 & 0.08 \\
			&0& 0.954 & 0.23 & 0.00 & 0.930 & 0.21 & 0.03 & 0.000 & 0.24 & 0.000 & 0.03 & 0.000 & 0.08 \\
			\hline
		\end{tabular}
	}
	\centering
	\caption{Average coverage probabilities (cov.) and average standard deviation (s.d.) for the 95\% confidence intervals computed via proposed MR-Local($\tau_0=1.6$), MR-Local+($\tau_0=1.6$), MR-MBE, TSHT, and MR-Raps in settings (a) to (e). The column ``$\emptyset$'' indicates the chance of occuring $\{\widehat{\mathcal{B}}=\emptyset\}$.}
	\label{tab-inf2}
\end{table}

\section{ Real studies based on two-sample GWAS summary data}
\label{sec-data}
In this section, we estimate the causal relationships based on the GWAS summary statistics from the OpenGWAS database \citep{elsworth2020mrc}. We infer the causal effect of
body mass index (BMI) on four other commonly studied traits: height (HGT), systolic blood pressure (SBP), type-2 diabetes (T2D), and coronary artery disease (CAD). To confirm the reliability of our proposal, we also estimate the causal effect of BMI on BMI, where the true value is one if certain assumptions are satisfied by two datasets. Specifically, we use the summary statistics for BMI from the GIANT consortium as the exposure and the summary statistics for BMI from the MRC-IEU consortium as the outcome.
Detailed information on the studies and the LD-clumping steps are given in the supplements.

\begin{table}[!htbp]
	\centering
	{\small
		\begin{tabular}{|c|c|c|c|c|c|}
			\hline
			Outcome   & MR-Local & MR-Local+ &
			MR-MBE& TSHT &MR-Raps\\
			\hline
			BMI  &  1.033(0.018)  &  1.033(0.017)   & 1.002(0.035)&1.000(0.006) & 1.010(0.007)\\
			\hline
			HGT & -0.055(0.012)* & -0.055(0.012)* & -0.041(0.030) &-0.033(0.004)  &-0.054(0.011)\\
			\hline
			SBP & 0.125(0.010)* & 0.125(0.010)* &0.126(0.048)&0.130(0.006)&0.120(0.010)\\
			\hline
			T2D & 0.008(0.001) & 0.008(0.001)   & 0.009(0.003) &0.007(0.000) &0.008 (0.001)\\
			\hline
			CAD & 0.378(0.093) & 0.378(0.061) & 0.308(0.177) &0.389(0.026)&0.366(0.029)\\
			\hline
		\end{tabular}
	}
	\caption{Causal inference of BMI on five traits: BMI, HGT, SBP, T2D, and CAD. Each column reports the estimated causal effects (estimated standard deviations). The results with ``*'' corresponds to $\widehat{\mathcal{B}}=\emptyset$.}
	\label{tab-dat}
\end{table}

The results are reported in Table \ref{tab-dat}. The estimates for the causal effect of BMI on BMI are close to one with 95\% confidence intervals covering the value one. Although the true causal effect of BMI on BMI should be one,  the actual causal effect based on these two populations may not be exactly one since the two datasets in use may be subject to different batch effects. For the causal effect of BMI on HGT, MR-Local and MR-Local+ yield estimates similar to MR-Raps, whereas the estimates based on the plurality rule show slight differences.   MR-Local and MR-Local+ algorithms output $\widehat{\mathcal{B}}=\emptyset$, suggesting that the results based on the balanced pleiotropy assumption are likely more reliable.
Regarding the causal effect estimates of BMI on CAD, all methods produce similar results, but MR-MBE has a significantly larger estimated standard deviation, resulting in a less efficient estimate.

\section{Discussion}
\label{sec-diss}
This work introduces MR-Local, a method that utilizes the local distribution to remove invalid IVs and then conduct causal inference. Our proposed causal estimate enjoys consistency and asymptotic normality under mild conditions and has reliable performance across various numerical scenarios. MR-Local is applicable under both the plurality rule and the balanced pleiotropy assumption. Hence, MR-Local is more robust than methods relying on these two assumptions in complex real-world situations. 


We discuss other statistics beyond the uncertainty measure defined in (\ref{Qhat-j}). While the variance entails second-order information on the local distribution, alternative ways can include higher-order information. On the other hand, A popular nonparametric statistic can be the KS-type statistic, which measures the distance between the empirical distribution and the true distribution. Formally,
\begin{align}
	\label{ks-stat}
	\widehat{\textup{KS}}(b)=\sup_{t\in[-\tau_0,\tau_0]} \left|\frac{1}{|\widehat{\C}(b)|}\sum_{k\in\widehat{\C}(b)}\mathbbm{1}(\hat{z}_k(b)\leq t)-\frac{\Phi(t)-\Phi(-\tau_0)}{\Phi(\tau_0)-\Phi(-\tau_0)}\right|.
\end{align}
If $\widehat{\textup{KS}}(b)$ is large, the IVs in the $j$-th clique are unlikely to follow the truncated standard normal distribution as in (\ref{local-dist}). Hence, one can also screen out a candidate value $b_j$ if $\widehat{\textup{KS}}(b_j)$ is beyond a proper threshold based on its limiting distribution. Future works involve relaxing the assumption on the distribution of the invalid effects and generalizing our findings to multivariate MR scope \citep{burgess2015multivariable,Sanderson2019}, which remains an open question.



\section*{Acknowledgement}
This work was supported by the National Natural Science Foundation of China (grant no. 12201630), the Fundamental Research Funds for the Central Universities, and the Research Funds of Renmin University of China.
 \bibliographystyle{chicago}
\bibliography{reference_IV}

\begin{thebibliography}{}

\bibitem[\protect\citeauthoryear{Basmann}{Basmann}{1957}]{basmann1957TSLS}
Basmann, R.~L. (1957).
\newblock A generalized classical method of linear estimation of coefficients
  in a structural equation.
\newblock {\em Econometrica: Journal of the Econometric Society\/}~{\em
  25\/}(1), 77--83.

\bibitem[\protect\citeauthoryear{Bowden, Davey~Smith, and Burgess}{Bowden
  et~al.}{2015}]{Bowden15}
Bowden, J., G.~Davey~Smith, and S.~Burgess (2015).
\newblock Mendelian randomization with invalid instruments: effect estimation
  and bias detection through egger regression.
\newblock {\em International Journal of Epidemiology\/}~{\em 44\/}(2),
  512--525.

\bibitem[\protect\citeauthoryear{Bowden, Davey~Smith, Haycock, and
  Burgess}{Bowden et~al.}{2016}]{Bowden16}
Bowden, J., G.~Davey~Smith, P.~C. Haycock, and S.~Burgess (2016).
\newblock Consistent estimation in mendelian randomization with some invalid
  instruments using a weighted median estimator.
\newblock {\em Genetic Epidemiology\/}~{\em 40\/}(4), 304--314.

\bibitem[\protect\citeauthoryear{Bowden, Del Greco~M, Minelli, Zhao, Lawlor,
  Sheehan, Thompson, and Davey~Smith}{Bowden et~al.}{2019}]{Bowden:2019aa}
Bowden, J., F.~Del Greco~M, C.~Minelli, Q.~Zhao, D.~A. Lawlor, N.~A. Sheehan,
  J.~Thompson, and G.~Davey~Smith (2019).
\newblock Improving the accuracy of two-sample summary-data mendelian
  randomization: moving beyond the nome assumption.
\newblock {\em International Journal of Epidemiology\/}~{\em 48\/}(3),
  728--742.

\bibitem[\protect\citeauthoryear{Burgess, Butterworth, and Thompson}{Burgess
  et~al.}{2013}]{Burgess:2013aa}
Burgess, S., A.~Butterworth, and S.~G. Thompson (2013).
\newblock Mendelian randomization analysis with multiple genetic variants using
  summarized data.
\newblock {\em Genetic Epidemiology\/}~{\em 37\/}(7), 658--665.

\bibitem[\protect\citeauthoryear{Burgess and Thompson}{Burgess and
  Thompson}{2015}]{burgess2015multivariable}
Burgess, S. and S.~G. Thompson (2015).
\newblock Multivariable mendelian randomization: the use of pleiotropic genetic
  variants to estimate causal effects.
\newblock {\em American Journal of Epidemiology\/}~{\em 181\/}(4), 251--260.

\bibitem[\protect\citeauthoryear{Cochran}{Cochran}{1954}]{cochran1954combination}
Cochran, W.~G. (1954).
\newblock The combination of estimates from different experiments.
\newblock {\em Biometrics\/}~{\em 10\/}(1), 101--129.

\bibitem[\protect\citeauthoryear{Davey~Smith and Ebrahim}{Davey~Smith and
  Ebrahim}{2003}]{Davey03}
Davey~Smith, G. and S.~Ebrahim (2003).
\newblock Mendelian randomization: can genetic epidemiology contribute to
  understanding environmental determinants of disease?
\newblock {\em International Journal of Epidemiology\/}~{\em 32\/}(1), 1--22.

\bibitem[\protect\citeauthoryear{Davey~Smith and Hemani}{Davey~Smith and
  Hemani}{2014}]{DaveySmith14}
Davey~Smith, G. and G.~Hemani (2014).
\newblock Mendelian randomization: genetic anchors for causal inference in
  epidemiological studies.
\newblock {\em Human Molecular Genetics\/}~{\em 23\/}(R1), R89--R98.

\bibitem[\protect\citeauthoryear{Elsworth, Lyon, Alexander, Liu, Matthews,
  Hallett, Bates, Palmer, Haberland, Smith, et~al.}{Elsworth
  et~al.}{2020}]{elsworth2020mrc}
Elsworth, B., M.~Lyon, T.~Alexander, Y.~Liu, P.~Matthews, J.~Hallett, P.~Bates,
  T.~Palmer, V.~Haberland, G.~D. Smith, et~al. (2020).
\newblock The mrc ieu opengwas data infrastructure.
\newblock {\em bioRxiv\/}.

\bibitem[\protect\citeauthoryear{Guo, Kang, Cai, and Small}{Guo
  et~al.}{2018}]{TSHT}
Guo, Z., H.~Kang, T.~T. Cai, and D.~S. Small (2018).
\newblock Confidence intervals for causal effects with invalid instruments by
  using two-stage hard thresholding with voting.
\newblock {\em Journal of the Royal Statistical Society Series B: Statistical
  Methodology\/}~{\em 80\/}(4), 793--815.

\bibitem[\protect\citeauthoryear{Hartwig, Davey~Smith, and Bowden}{Hartwig
  et~al.}{2017}]{Hartwig:2017aa}
Hartwig, F.~P., G.~Davey~Smith, and J.~Bowden (2017).
\newblock Robust inference in summary data mendelian randomization via the zero
  modal pleiotropy assumption.
\newblock {\em International Journal of Epidemiology\/}~{\em 46\/}(6),
  1985--1998.

\bibitem[\protect\citeauthoryear{Holland}{Holland}{1988}]{holland1988causal}
Holland, P.~W. (1988).
\newblock Causal inference, path analysis and recursive structural equations
  models.
\newblock {\em ETS Research Report Series\/}~{\em 1988\/}(1), i--50.

\bibitem[\protect\citeauthoryear{Hu, Zhao, Lin, Wang, Peng, Zhao, Wan, and
  Yang}{Hu et~al.}{2022}]{hu2022mendelian}
Hu, X., J.~Zhao, Z.~Lin, Y.~Wang, H.~Peng, H.~Zhao, X.~Wan, and C.~Yang (2022).
\newblock Mendelian randomization for causal inference accounting for
  pleiotropy and sample structure using genome-wide summary statistics.
\newblock {\em Proceedings of the National Academy of Sciences\/}~{\em
  119\/}(28), e2106858119.

\bibitem[\protect\citeauthoryear{Kang, Zhang, Cai, and Small}{Kang
  et~al.}{2016}]{Kang16}
Kang, H., A.~Zhang, T.~T. Cai, and D.~S. Small (2016).
\newblock Instrumental variables estimation with some invalid instruments and
  its application to mendelian randomization.
\newblock {\em Journal of the American Statistical Association\/}~{\em
  111\/}(513), 132--144.

\bibitem[\protect\citeauthoryear{Koles{\'a}r, Chetty, Friedman, Glaeser, and
  Imbens}{Koles{\'a}r et~al.}{2015}]{Kole15}
Koles{\'a}r, M., R.~Chetty, J.~Friedman, E.~Glaeser, and G.~W. Imbens (2015).
\newblock Identification and inference with many invalid instruments.
\newblock {\em Journal of Business \& Economic Statistics\/}~{\em 33\/}(4),
  474--484.

\bibitem[\protect\citeauthoryear{Morrison, Knoblauch, Marcus, Stephens, and
  He}{Morrison et~al.}{2020}]{morrison2020ng}
Morrison, J., N.~Knoblauch, J.~H. Marcus, M.~Stephens, and X.~He (2020).
\newblock Mendelian randomization accounting for correlated and uncorrelated
  pleiotropic effects using genome-wide summary statistics.
\newblock {\em Nature Genetics\/}~{\em 52\/}(7), 740–747.

\bibitem[\protect\citeauthoryear{Qi and Chatterjee}{Qi and
  Chatterjee}{2019}]{qi2019mendelian}
Qi, G. and N.~Chatterjee (2019).
\newblock Mendelian randomization analysis using mixture models for robust and
  efficient estimation of causal effects.
\newblock {\em Nature Communications\/}~{\em 10\/}(1), 1--10.

\bibitem[\protect\citeauthoryear{Sanderson, {Davey Smith}, Windmeijer, and
  Bowden}{Sanderson et~al.}{2019}]{Sanderson2019}
Sanderson, E., G.~{Davey Smith}, F.~Windmeijer, and J.~Bowden (2019).
\newblock {An examination of multivariable Mendelian randomization in the
  single-sample and two-sample summary data settings}.
\newblock {\em International Journal of Epidemiology\/}~{\em 48\/}(3),
  713--727.

\bibitem[\protect\citeauthoryear{Small}{Small}{2007}]{small2007sensitivity}
Small, D.~S. (2007).
\newblock Sensitivity analysis for instrumental variables regression with
  overidentifying restrictions.
\newblock {\em Journal of the American Statistical Association\/}~{\em
  102\/}(479), 1049--1058.

\bibitem[\protect\citeauthoryear{Sun, Cui, and Tchetgen}{Sun
  et~al.}{2022}]{sun2022selective}
Sun, B., Y.~Cui, and E.~T. Tchetgen (2022).
\newblock Selective machine learning of the average treatment effect with an
  invalid instrumental variable.
\newblock {\em The Journal of Machine Learning Research\/}~{\em 23\/}(1),
  9249--9288.

\bibitem[\protect\citeauthoryear{Verbanck, Chen, Neale, and Do}{Verbanck
  et~al.}{2018}]{Verbanck:2018aa}
Verbanck, M., C.-Y. Chen, B.~Neale, and R.~Do (2018).
\newblock Detection of widespread horizontal pleiotropy in causal relationships
  inferred from mendelian randomization between complex traits and diseases.
\newblock {\em Nature Genetics\/}~{\em 50\/}(5), 693--698.

\bibitem[\protect\citeauthoryear{Windmeijer, Farbmacher, Davies, and
  Davey~Smith}{Windmeijer et~al.}{2019}]{Windmeijer2019}
Windmeijer, F., H.~Farbmacher, N.~Davies, and G.~Davey~Smith (2019).
\newblock On the use of the lasso for instrumental variables estimation with
  some invalid instruments.
\newblock {\em Journal of the American Statistical Association\/}~{\em
  114\/}(527), 1339--1350.

\bibitem[\protect\citeauthoryear{Windmeijer, Liang, Hartwig, and
  Bowden}{Windmeijer et~al.}{2021}]{Wind21}
Windmeijer, F., X.~Liang, F.~P. Hartwig, and J.~Bowden (2021).
\newblock The confidence interval method for selecting valid instrumental
  variables.
\newblock {\em Journal of the Royal Statistical Society Series B: Statistical
  Methodology\/}~{\em 83\/}(4), 752--776.

\bibitem[\protect\citeauthoryear{Ye, Shao, and Kang}{Ye
  et~al.}{2021}]{ye2021debiased}
Ye, T., J.~Shao, and H.~Kang (2021).
\newblock Debiased inverse-variance weighted estimator in two-sample
  summary-data mendelian randomization.
\newblock {\em The Annals of Statistics\/}~{\em 49\/}(4), 2079--2100.

\bibitem[\protect\citeauthoryear{Zhao}{Zhao}{2018}]{mrraps}
Zhao, Q. (2018).
\newblock {\em mr.raps: Two Sample Mendelian Randomization using Robust
  Adjusted Profile Score}.
\newblock R package version 0.2.

\bibitem[\protect\citeauthoryear{Zhao, Wang, Hemani, Bowden, and Small}{Zhao
  et~al.}{2020}]{zhao2018statistical}
Zhao, Q., J.~Wang, G.~Hemani, J.~Bowden, and D.~S. Small (2020).
\newblock Statistical inference in two-sample summary-data mendelian
  randomization using robust adjusted profile score.
\newblock {\em The Annals of Statistics\/}~{\em 48\/}(3), 1742--1769.

\end{thebibliography}

\end{document}